\documentclass[showpacs, twocolumn,prl,notitlepage]{revtex4}
\usepackage{amssymb}
\usepackage{amsmath}
\usepackage{graphicx}
\usepackage{bm}
\usepackage{hyperref}
\usepackage{subfigure}
\usepackage{soul}
\usepackage{color}
\usepackage{dcolumn}

\begin{document}

\title{Effect of initial-state nucleon-nucleon correlations on
collective flow in ultra-central heavy-ion collisions}
\author{G.\ S.\ Denicol${}^{a}$, C.\ Gale${}^{a, b}$, S.\ Jeon${}^{a}$, J.-F.~Paquet${}^{a}$, and B.\
Schenke${}^{c}$}
\affiliation{$^{a}$Department of Physics, McGill University, 3600 University Street,
Montreal, Quebec, H3A 2T8, Canada}
\affiliation{$^{b}$Frankfurt Institute for Advanced Studies, Ruth-Moufang-Str. 1, D-60438
Frankfurt am Main, Germany}
\affiliation{$^{c}$Physics Department, Bldg. 510A, Brookhaven National Laboratory, Upton,
NY-11973, USA}

\begin{abstract}
We investigate the effect of nucleon-nucleon correlations on the initial
condition of ultra-central heavy ion collisions at LHC energies. We
calculate the eccentricities of the MC-Glauber and IP-Glasma models in the
0--1\% centrality class and show that they are considerably affected by the
inclusion of such type of correlations. For an IP-Glasma initial condition,
we further demonstrate that this effect survives the fluid-dynamical
evolution of the system and can be observed in its final state azimuthal
momentum anisotropy.
\end{abstract}

\maketitle


\section{Introduction}

Various observations in ultra-relativistic heavy-ion collisions at the
energies available at the Relativistic Heavy Ion Collider (RHIC) and the
Large Hadron Collider (LHC) indicate that a quark-gluon-plasma (QGP) near
thermodynamic equilibrium is created \cite{Gale:2013da}. One of the most surprising results
obtained at the RHIC, and later confirmed at the LHC, is that this novel
state of nuclear matter behaves as an almost perfect fluid, with
one of the smallest shear viscosity to entropy density ratios, $\eta /s$,
ever observed in nature \cite%
{whitepaper1,whitepaper2,whitepaper3,whitepaper4,LHC}. Currently, extracting the
shear viscosity of the QGP from experiment is one of the main challenges in
high energy nuclear physics.

The initial energy density profile of the QGP is one of the main ingredients
in the fluid-dynamical simulation of a heavy ion collision. It is also one
of the main sources of uncertainties when attempting to extract the
properties of the QGP by fitting heavy ion collision measurements \cite%
{Luzum:2012wu,Song:2008hj}. There are several models used to provide this
input, from the Monte-Carlo Glauber models \cite{Miller:2007ri} to
saturation based models, such as the Monte-Carlo KLN \cite{KLN}, the rcBK 
\cite{Albacete:2012xq,Dumitru:2012yr}, or the IP-Glasma \cite%
{Schenke:2012wb,Schenke:2012fw} model. While these models differ in many
aspects, they share one unique ingredient: they all require, as an input,
the nucleon configurations of each colliding nucleus in an event. The
fluctuation of such nucleon configurations on an event by event basis is a
quantum effect and is the dominant contribution to the fluctuation of the
azimuthal momentum distribution of charged hadrons observed at the RHIC and
the LHC.

Practically all current initial condition models assume that the nucleons
inside each colliding heavy ion are independent point-like particles, with
positions prescribed according to a Woods-Saxon charge distribution
function. This is, of course, a very crude approximation since it is well
known that the nucleons inside nuclei are not independent from each other
and should exhibit some degree of correlation. The question that follows is
what kind of effect such nucleon-nucleon correlations can have on the
initial state of a heavy ion collision and whether the effect is strong
enough to leave its imprints on the final state of the system.

In Ref.~\cite{Alvioli:2009ab} Alvioli \textit{et al} determined nucleon
profiles of heavy nuclei that take into account the effect of
nucleon-nucleon correlations. The phenomenological implications of such
correlations on the initial state of heavy ion collisions were then
investigated in Ref.~\cite{Alvioli:2011sk}. Such studies were implemented
for central and peripheral heavy ion collisions and indicated that the
effect of correlations on the initial collision geometry was rather small.
Additional studies on the effect of nucleon-nucleon correlations on the
initial state of heavy ion collisions were also conducted in Refs.~\cite%
{All1,All} and similar conclusions were reached.

Recently, the CMS, ATLAS, and ALICE collaborations were able to isolate 
\textit{ultra-central} heavy ion collisions \cite%
{CMS:2013bza,ATLAS:2012at,ALICE:2011ab} at the LHC, i.e., rare heavy ion
collisions which have extremely large multiplicity and, consequently, a very
large system size. By analyzing the collective flow signals of only such
rare events, it is believed that an even more direct observation of the
hydrodynamic behavior of QCD matter would be possible, allowing more
accurate extractions of the viscous properties of the QGP \cite{Luzum:2012wu}%
. However, one first has to determine whether a precise fluid-dynamical
simulation of ultra-central collisions requires the inclusion of
nucleon-nucleon correlations.

In this paper, we investigate for the first time the effect that
nucleon-nucleon correlations can have on the initial condition of
ultra-central heavy ion collisions at LHC energies. We show that the
eccentricities of the MC-Glauber and IP-Glasma models in the 0--1\%
centrality class are considerably affected by the inclusion of such type of
correlations. For an IP-Glasma initial condition, we further demonstrate
that this effect survives the fluid-dynamical evolution of the system and
can be observed in its final state azimuthal momentum anisotropy. The
inclusion of correlations improves the agreement with experimental data
noticeably, even though more work remains necessary in order to obtain a
satisfactory agreement.

\section{Effect of nucleon-nucleon correlations on eccentricities}

A key ingredient present in all initial state models is the configuration of
the nucleons inside each nucleus participating in the collision. The
position of the protons and neutrons inside the nucleus are usually sampled
independently according to a Woods-Saxon distribution,%
\begin{equation}
\rho \left( r\right) =\frac{\rho _{0}}{1+\exp \left( \frac{r-R}{a}\right) }.
\end{equation}%
For a Pb nucleus, the parameters $R$ and $a$ are given by $R=6.62$ fm and $%
a= $ $0.546$ fm. The constant $\rho _{0}$ is the nucleon density and is not
important when sampling the position of each nucleon since the distribution
above is renormalized so that its integral in $r$ becomes one. In this
procedure all sources of nucleon-nucleon correlations are neglected.

Recently, Alvioli, Drescher and Strikman presented a Monte-Carlo generator
of nucleon configurations that takes into account several sources of
nucleon-nucleon correlations \cite{Alvioli:2009ab}. In their approach, the
nucleus is described by a multiparticle wavefunction that depends on the
position of all nucleons, as well as their spin and isospin. The position of
each proton and neutron that constitutes the Pb nuclei is then found
iteratively using the squared amplitude of the wave function \cite%
{Alvioli:2005cz} together with the constraint that, for every configuration,
the single nucleon distribution function remains equal to the Woods-Saxon
distribution shown above. Such configurations provide more realistic
two-nucleon distribution functions than the ones obtained by sampling the
nucleons independently, even if the excluded volume approximation is used in
the latter procedure. In this work we study the effect of correlations by
employing the Pb nuclei sampled by Alvioli \textit{et al} using the
formalism described in \cite{Alvioli:2009ab}, which were made public and can
be found in \cite{website}.

We note that, due to numerical limitations, the nuclei configurations
presented in Ref.~\cite{Alvioli:2009ab} did not take into account all the
possible sources of correlations, i.e., only effective central correlations
were introduced. Nevertheless, the aforementioned calculation should be able
to capture the basic features of the 2-nucleon correlation functions and
should provide a good insight when estimating the effect of correlations in
ultra-central collisions. We note that more realistic configurations of
nuclei were later calculated in Ref.~\cite{Alvioli:2011sk}, but only for Au
nuclei and not for Pb.

In the Monte-Carlo Glauber model, the positions of the protons and neutrons
of each heavy ion involved in the collision are used to calculate the
position of the wounded nucleons and of binary collisions in the transverse
plane of the reaction. The initial entropy density (or energy density) of
the system, $s$ (or $\varepsilon $), is then assumed to scale as a linear
combination of the density of binary collisions, $n_{BC}$, and the density
of wounded nucleons, $n_{WN}$. That is, $s\sim fn_{BC}+\left( 1-f\right)
n_{WN}$, with $f$ being the fraction of binary collisions. The main inputs
required by the MC-Glauber model are the impact parameter of the collision,
the corresponding nucleon-nucleon inelastic cross-section, $\sigma _{NN}$,
and the probability distribution of a binary collision or wounded nucleon to
create a certain amount of entropy, $S$. The overall normalization of the
initial entropy density profile and the fraction of binary collisions are
free parameters of the model, that should be determined so that the
multiplicity distribution of the system is well reproduced. Here, we shall
use $f=0.14$, $\sigma _{NN}=62$ mb, and an entropy distribution per binary
collision/wounded nucleon given by a negative binomial distribution,%
\begin{equation}
P(S)=\frac{\Gamma (S+\kappa )}{\Gamma (\kappa )\Gamma (S+1)}\frac{r^{S}}{%
(r+1)^{S+\kappa }},
\end{equation}
with $r=18.4$ and $\kappa =1.25$. These parameters provide a reasonable
description of the multiplicity distribution of heavy ion collisions at the
LHC.

On the other hand, in the IP-Glasma model the positions of protons and
neutrons inside the nuclei are used to reconstruct the color charge density
of the heavy ions colliding. This information is then used in the model to
calculate the energy momentum tensor of the system at the initial stages of
the collision, by solving the classical Yang-Mills equations. This model
requires as an input the impact parameter of the collision and the
thermalization time. The energy dependence of the local saturation scale,
which affects the magnitude of the energy density, is given by the IP-Sat
dipole model~\cite{Kowalski:2003hm}, with its parameters fit to deeply
inelastic scattering data \cite{Rezaeian:2012ji}. In the IP-Glasma model,
negative binomial distributions appear naturally \cite{Schenke:2013dpa}. For
details of this model, see Refs.~\cite{Schenke:2012wb,Schenke:2012fw}.

We first check the effect of correlations on the root mean square
eccentricities $\varepsilon _{n}\{2\}$,%
\begin{equation}
\varepsilon _{n}\{2\}\equiv \sqrt{\left\langle e_{n}e_{n}^{\ast
}\right\rangle _{\mathrm{events}}}\text{, \ }e_{n}=\left\langle r^{n}\exp
\left( in\phi \right) \right\rangle _{\varepsilon ,s}\text{ },
\end{equation}%
of the initial state profile of the MC-Glauber and IP-Glasma models. Above,
the brackets denote the spatial integrals (performed in a given event), $%
\left\langle \cdots \right\rangle _{A}=\int rdrd\phi $ $\left( \cdots
\right) $ $A\left( r,\phi ,\tau _{0}\right) $, with$\ r$ and $\phi $ being
the coordinates in the transverse plane, $\tau _{0}$ the thermalization
time, and $A$ is either the entropy density, $s$, or the energy density, $%
\varepsilon $. The eccentricities characterize the transverse shape of the
fluid at the thermalization time $\tau _{0}$ and influence directly the
magnitude of the anisotropic flow produced during the fluid-dynamical
evolution of the system \cite{Teaney:2012ke,Niemi:2012aj}.

\begin{figure}[th]
\subfigure[]{\includegraphics[width=0.40\textwidth]{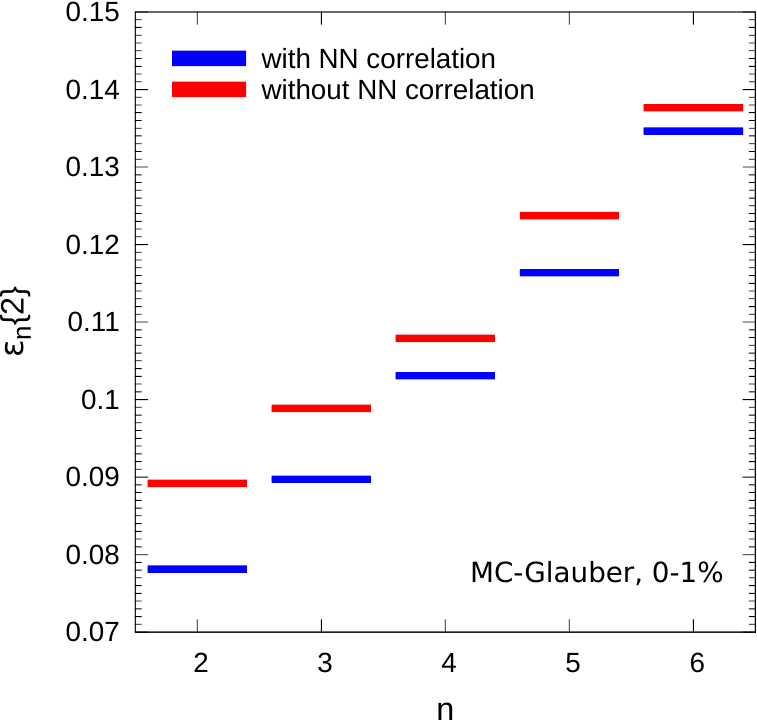}
\label{figEn:subfigure2}} 
\subfigure[]{\includegraphics[width=0.40\textwidth]{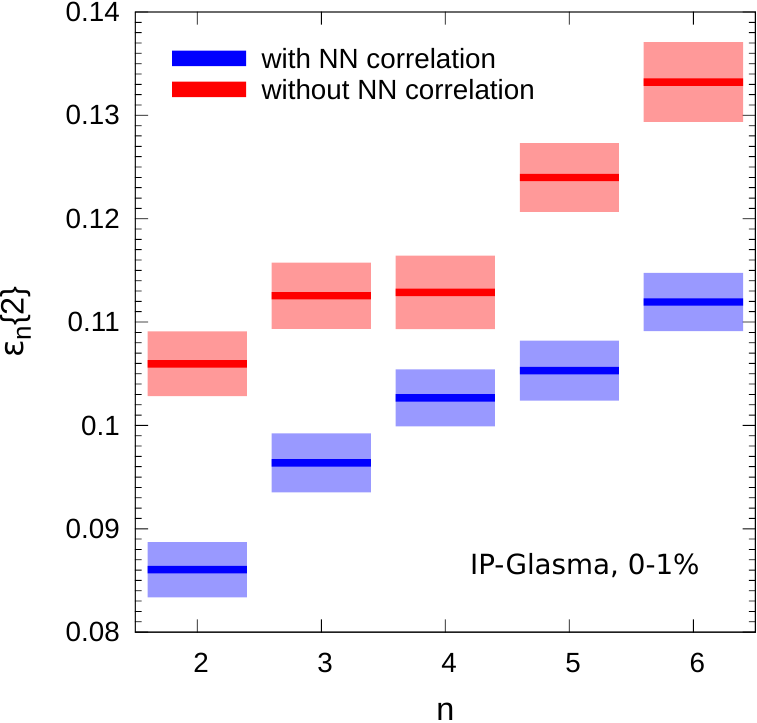}
\label{figEn:subfigure3}}
\caption{(Color online) Comparison of the eccentricities, $\varepsilon%
_{n}{\{2\}}$, of the MC-Glauber model (top panel) and IP-Glasma model
(bottom panel) with (blue) and without (red) the effect of nucleon-nucleon
correlations. The results shown are for the 0-1\% most central events. The
statistical uncertainty of the calculation is given by the bands. }
\label{Comparison}
\end{figure}

In Fig.~\ref{Comparison} we show the root mean square eccentricities $%
\varepsilon _{n}\left\{ 2\right\} $, for $n=2$--$6$, for Pb-Pb collisions in
the 0-1\% centrality class. We note that the bracket $\left\langle \cdots
\right\rangle _{\mathrm{events}}$ denotes an average over events. We present
results for the MC-Glauber model (top panel) and the IP-Glasma model (bottom
panel). In this work, the eccentricities of the Glauber model were computed
using the entropy density as weights while those of the IP-Glasma model were
computed using energy density as weights. The blue lines correspond to the
results with nucleon-nucleon correlations while the red lines to the results
without such correlations. The effect of correlations is the largest for
IP-Glasma initial conditions, where $\varepsilon _{2}$ changes by almost $%
20\%$ with the inclusion of nucleon-nucleon correlations. For the MC-Glauber
model considered the effect of correlations on the eccentricities becomes
smaller, with $\varepsilon _{2}$ being reduced by 10\% with the inclusion of
correlations. We note that the effect of nucleon-nucleon correlations on the
MC-Glauber model depends on the fraction of binary collisions. If we had
taken $f=1$, the changes in eccentricity would become considerably larger,
even though not as large as those occurring for IP-Glasma initial
conditions. However, when $f=1$ one cannot describe the multiplicity
distribution of heavy ion collisions very well, so we refrain from including
such results in this paper.

We checked that, for non-central collisions in the 20-30\% centrality class,
the effect of correlations on the eccentricities become very small for both
initial condition models.

In the following, we shall demonstrate that such changes in the
eccentricities of the IP-Glasma initial condition can lead to large changes
on the transverse momentum anisotropy of the final state of the collision.

\section{Fluid-dynamical simulation of ultra-central collision}

In order to calculate the flow harmonics, one must compute the
fluid-dynamical evolution of the bulk nuclear matter created in the
collision. In the following study the initial state is given by the
IP-Glasma model, in the 0-1\% and 2.5-5\% centrality classes, with a
thermalization time of $\tau _{0}=0.4$ fm. The fluid-dynamical evolution of
the system with this initial state is solved numerically using the \textsc{%
music} simulation \cite{Schenke:2010nt,Marrochio:2013wla}. The main
equations of motion solved by \textsc{music} are the conservation laws of
energy and momentum,%
\begin{equation}
\partial _{\mu }T^{\mu \nu }=0,
\end{equation}%
where $T^{\mu \nu }=(\varepsilon +P)u^{\mu }u^{\nu }-g^{\mu \nu }P+\pi ^{\mu
\nu }$, with $\varepsilon $, $P$, $u^{\mu }$, and $\pi ^{\mu \nu }$ being
the energy density, thermodynamic pressure, 4-velocity, and shear-stress
tensor, respectively. The time evolution of $\pi ^{\mu \nu }$ is obtained by
solving a version of Israel-Stewart theory \cite{IS,DNMR},%
\begin{equation}
\tau _{\pi }\dot{\pi}^{\left\langle \mu \nu \right\rangle }+\pi ^{\mu \nu
}=2\eta \sigma ^{\mu \nu }-\frac{4}{3}\tau _{\pi }\pi ^{\mu \nu }\partial
_{\mu }u^{\mu },
\end{equation}%
where $\dot{\pi}^{\left\langle \mu \nu \right\rangle }\equiv \Delta _{\alpha
}^{\mu }\Delta _{\beta }^{\nu }u^{\lambda }\partial _{\lambda }\pi ^{\alpha
\beta }$, with $\Delta _{\alpha }^{\mu }=g_{\alpha }^{\mu }-u^{\mu
}u_{\alpha }$ being the projection operator orthogonal to the flow velocity.
In our calculations the flow velocity is defined according to the Landau
picture \cite{Landau}, i.e., $T^{\mu \nu }u_{\mu }=\varepsilon u^{\nu }$.
The equation of state, $P\left( \varepsilon \right) $, employed in our
simulation is the\ s95p-PCE-v1 parametrization of lattice QCD calculations 
\cite{Huovinen:2009yb}, with a chemical freeze-out temperature of $T_{%
\mathrm{chem}}=150$ MeV. The shear viscosity to entropy density ratio is set
to $\eta /s=0.22$ and the relaxation time is given by $\tau _{\pi }=5\eta
/\left( \varepsilon +P\right) $. Finally, we used a freezeout temperature of 
$T=103$ MeV. We note that these parameters are not exactly the same as those
employed in Ref.~\cite{Gale:2012rq}. Nevertheless, we checked that they are
able to provide an equally good description of semi-peripheral and
peripheral heavy ion collisions at LHC energies \cite{QM14}. We note that
the value of $\eta/s$ employed in our calculation is slightly higher than
the one used in Ref.~\cite{Gale:2012rq}, but this is mainly due to the
difference in the freezeout temperatures employed in each calculations.

We compute the root mean square of the flow harmonic coefficients, $V_{n}$,%
\begin{equation}
v_{n}\{2\}=\sqrt{\left\langle V_{n}V_{n}^{\ast }\right\rangle _{\mathrm{%
events}}},\text{ \ }V_{n}=\left\langle \exp \left( in\phi \right)
\right\rangle ,
\end{equation}%
where $\phi $ is the azimuthal transverse momentum angle and the bracket
denotes the following averaging procedure (performed in a given event) 
\begin{equation}
\left\langle \cdots \right\rangle =\frac{\int d\phi d^{2}p_{T}\text{ }%
f\left( p_{T},\phi \right) \times \left( \cdots \right) }{\int d\phi
d^{2}p_{T}\text{ }f\left( p_{T},\phi \right) },
\end{equation}%
with $f_{p}$ being the single particle probability distribution function
obtained from the Cooper-Frye formalism \cite{Cooper:1974mv}. As before, the
bracket $\left\langle \cdots \right\rangle _{\mathrm{events}}$ denotes an
average over events. The nonequilibrium momentum distribution function of
hadrons, $f^{i}\left( k\right) $, included in the Cooper-Frye description is
given by the usual 14-moment approximation \cite{IS},%
\begin{equation*}
f^{i}\left( k\right) =f_{0}^{i}\left( k\right) \left[ 1+\tilde{f}%
_{0}^{i}\left( k\right) \frac{\pi ^{\mu \nu }k_{i}^{\mu }k_{i}^{\nu }}{%
2\left( \varepsilon +P\right) T^{2}}\right] ,
\end{equation*}%
where the index $i$ refers to the corresponding hadron species, $k_{i}^{\mu
} $ is the 4-momentum of the $i$--th species, $f_{0}^{i}\left( k\right) $ is
the Bose-Einstein/Fermi-Dirac distribution, and $\tilde{f}_{0}^{i}\left(
k\right) =1-af_{0}^{i}\left( k\right) $, with $a=1(-1)$ for
fermions(bosons). The energy density, pressure, temperature, and
shear-stress tensor in this expression are those in the freeze-out surface
elements, that we sum over to get the particle spectra. In this calculation,
we include all hadrons/resonances with masses up to 1.3 GeV and consider all
2-body and three-body decays of unstable resonances.

\begin{figure}[th]
\centering
\subfigure[]{\includegraphics[width=0.40\textwidth]{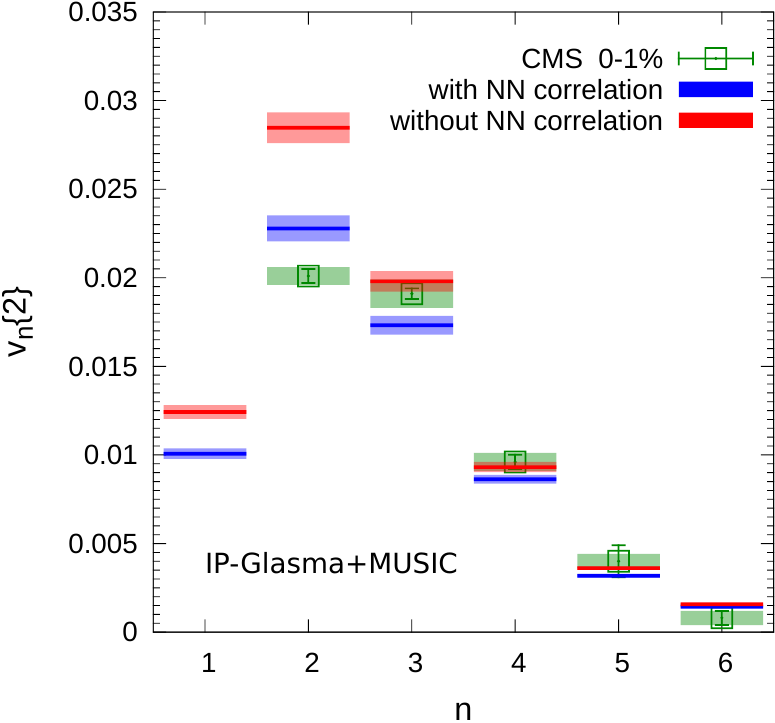}
\label{VnCMS}} \hspace{1.5cm} 
\subfigure[]{\includegraphics[width=0.40\textwidth]{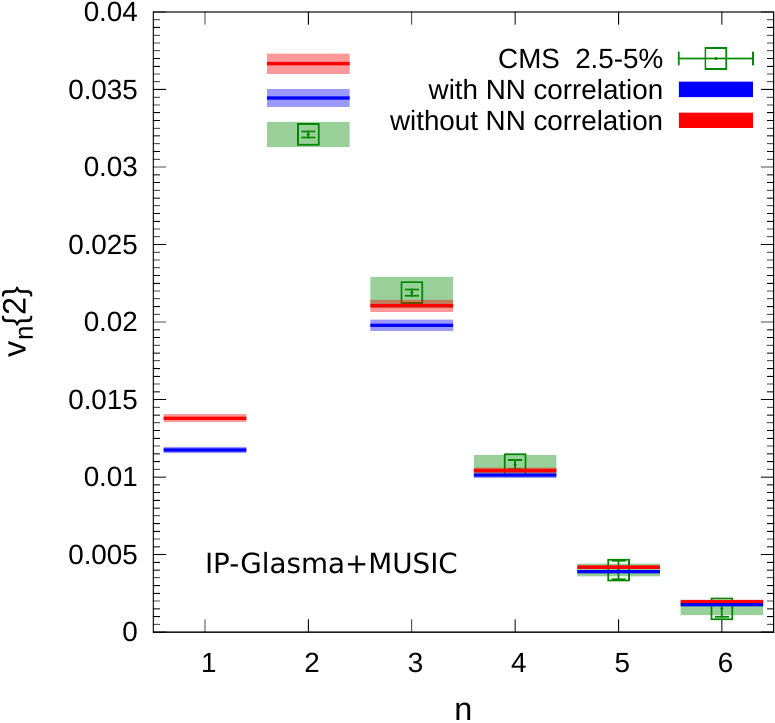}
\label{VnCMS2}}
\caption{ (Color online) Comparison of the flow harmonics, $v_n\{2\} $,
calculated using IP-Glasma initial conditions with (blue bars) and without
(red bars) the effect of correlations, with CMS data (green squares) for the
0--1 \% (top panel) and 2.5--5 \% (bottom panel) centrality classes. The statistical
uncertainty of the calculation is given by the bands. For the experimental
points, the bands denote systematic uncertainties while the bars denote the
statistical uncertainty.}
\end{figure}

The flow harmonics calculated with and without the effect of correlations
are compared to CMS data \cite{CMS:2013bza} in Fig.\ref{VnCMS} for the
0--1\% centrality class. In order to compare with CMS measurements, we
introduced a lower cut-off of 0.3 GeV on the transverse momentum integrals.
We see a very large effect of nucleon-nucleon correlations on the elliptic
and triangular flow of ultra-central collisions. As expected, the effect is
rather similar to the one encountered for the initial state eccentricities.
We note that the introduction of correlations produces a large reduction of
the elliptic flow coefficient, leading to a better agreement with the CMS
data.

For the sake of completeness, we also check the effect of correlations in
the 2.5--5\% centrality class, which would be usually referred to as central
collisions. In Fig.\ref{VnCMS2}, we compare flow harmonics calculated with
and without the effect of correlations to CMS data in this centrality class.
We still observe some effect of correlations on the elliptic and triangular
flow of the system. Nevertheless, the effect is considerably smaller than
the one observed in the ultra-central collisions.

We remark that the agreement between theory and experiment is still not
satisfactory in ultra-central collisions. While inclusion of nucleon-nucleon
correlations did improve the overall agreement between the flow harmonics
measured by CMS and those calculated from fluid-dynamical simulations, the
values of elliptic flow and triangular predicted from theory are still not in
agreement with the values observed by the experiments: The elliptic flow
coefficient calculated with the fluid-dynamical simulation is above the data
while the triangular flow coefficient is slightly below the data. The same
would be true if we had compared to data from the ATLAS or ALICE
experiments, which differ from CMS data by having different transverse
momentum cuts. 

Such partial disagreement with the data can happen because the effects of
nucleon-nucleon correlations were only partially included, e.g., the effect
of short-range attractive interactions and of 3-nucleon correlations were
not taken into account. Previous results indicate that the inclusion of
attractive interactions may reduce the effect of central correlations on the
initial state eccentricities. On the other hand, this has never been checked
for IP-Glasma initial conditions and will be an interesting project for the
future. It is also possible that such disagreement with the data may
originate from the fluid-dynamical simulation of the bulk QCD matter itself.
Our calculation did not take into account some sources of dissipation such
as bulk viscosity and also did not include higher-order non-linear terms
that appear in the fluid-dynamical equations of motion \cite%
{Denicol:2014vaa}. We shall discuss these two ingredients in future
publications.

\section{Conclusions}

In this paper, we calculated the effect of nucleon-nucleon correlations on
the initial state of ultra-central heavy ion collisions. We showed that
these correlations have a large effect on the eccentricities of the
MC-Glauber and IP-Glasma models, leading to significant reduction of $%
\varepsilon _{2}$ and $\varepsilon _{3}$. For IP-Glasma initial conditions,
we further calculated the final state momentum anisotropy coefficients, $%
v_{n}\{2\}$, and demonstrated that they are also considerably affected by
the initial state nucleon-nucleon correlations. Overall, the inclusion of
nucleon-nucleon correlations improves the agreement with experimental
measurements of flow coefficients and is a first step towards achieving a
reasonable description of such rare events. This indicates that
ultra-central collisions might be ideal, not only to investigate the bulk
properties of hot and dense nuclear matter, but also to study correlations
of the initial state wave function, thus providing a fascinating connection
between the physics of relativistic heavy-ion collisions and that of nuclear
structure.

\section{Acknowledgments}

The authors thank H.~Niemi, A.~Dumitru, M.~Luzum, and J.~B.~Rose for fruitful discussions. This
work was supported in part by the Natural Sciences and Engineering Research
Council of Canada, and by the U. S. DOE Contract No. DE-AC02-98CH10886.
G.S.~Denicol acknowledges support through a Banting Fellowship of the
Natural Sciences and Engineering Research Council of Canada, and C.~G.~
acknowledges support from the Hessian Initiative for Excellence (LOEWE)
through the Helmholtz International Center for FAIR (HIC for FAIR).


\begin{thebibliography}{99}

\bibitem{Gale:2013da} See, for instance, C.~Gale, S.~Jeon and B.~Schenke, 
Int.\ J.\ Mod.\ Phys.\ A \textbf{28}, 1340011 (2013), and references therein. 

\bibitem{whitepaper1} BRAHMS collaboration, Nuclear Physics A \textbf{757}
1, 1 (2005).

\bibitem{whitepaper2} PHENIX collaboration, Nuclear Physics A \textbf{757}
1, 184 (2005).

\bibitem{whitepaper3} PHOBOS collaboration, Nuclear Physics A \textbf{757}
1, 28 (2005).

\bibitem{whitepaper4} STAR collaboration, Nuclear Physics A \textbf{757} 1,
102 (2005).

\bibitem{LHC} See, for instance, Y.~Schutz and U.~Wiedemann, J. Phys. G: Nucl. Part. Phys. 38, 120301 (2011), and
all papers in this volume.

\bibitem{Luzum:2012wu} M.~Luzum and J.~-Y.~Ollitrault, 
Nucl.\ Phys.\ A \textbf{904-905}, 377c (2013). 

\bibitem{Song:2008hj} H.~Song and U.~W.~Heinz, 
J.\ Phys.\ G \textbf{36}, 064033 (2009) [arXiv:0812.4274 [nucl-th]]. 

\bibitem{Miller:2007ri} M.~L.~Miller, K.~Reygers, S.~J.~Sanders and
P.~Steinberg, 
Ann.\ Rev.\ Nucl.\ Part.\ Sci.\ \textbf{57}, 205 (2007) [nucl-ex/0701025]. 

\bibitem{KLN} D.~Kharzeev, E.~Levin and M.~Nardi, Phys. Rev. \textbf{C} 71,
054903 (2005); D. Kharzeev, E. Levin and M. Nardi, Nucl.~Phys.~\textbf{A}
747, 609 (2005).

\bibitem{Albacete:2012xq} J.~L.~Albacete, A.~Dumitru, H.~Fujii and Y.~Nara, 
Nucl.\ Phys.\ A \textbf{897}, 1 (2013). 


\bibitem{Dumitru:2012yr} A.~Dumitru and Y.~Nara, 
Phys.\ Rev.\ C \textbf{85}, 034907 (2012). 


\bibitem{Schenke:2012wb} B.~Schenke, P.~Tribedy and R.~Venugopalan, 
Phys.\ Rev.\ Lett.\ \textbf{108}, 252301 (2012) [arXiv:1202.6646 [nucl-th]]. 


\bibitem{Schenke:2012fw} B.~Schenke, P.~Tribedy and R.~Venugopalan, 
Phys.\ Rev.\ C \textbf{86}, 034908 (2012) [arXiv:1206.6805 [hep-ph]]. 


\bibitem{Alvioli:2009ab} M.~Alvioli, H.~-J.~Drescher and M.~Strikman, 
Phys.\ Lett.\ B \textbf{680}, 225 (2009) [arXiv:0905.2670 [nucl-th]]. 


\bibitem{Alvioli:2011sk} M.~Alvioli, H.~Holopainen, K.~J.~Eskola and
M.~Strikman, 
Phys.\ Rev.\ C \textbf{85}, 034902 (2012) [arXiv:1112.5306 [hep-ph]]. 

\bibitem{All1} J.~-P.~Blaizot, W.~Broniowski and J.~-Y.~Ollitrault, 
arXiv:1405.3274 [nucl-th]. 

\bibitem{All} W.~Broniowski and M.~Rybczynski, 
Phys.\ Rev.\ C \textbf{81}, 064909 (2010). 


\bibitem{CMS:2013bza} S.~Chatrchyan \textit{et al.} [CMS Collaboration], 
JHEP \textbf{1402}, 088 (2014) [arXiv:1312.1845 [nucl-ex]]. 


\bibitem{ATLAS:2012at} G.~Aad \textit{et al.} [ATLAS Collaboration], 
Phys.\ Rev.\ C \textbf{86}, 014907 (2012) [arXiv:1203.3087 [hep-ex]]. 


\bibitem{ALICE:2011ab} K.~Aamodt \textit{et al.} [ALICE Collaboration], 
Phys.\ Rev.\ Lett.\ \textbf{107}, 032301 (2011). 


\bibitem{Alvioli:2005cz} M.~Alvioli, C.~Ciofi degli Atti and H.~Morita, 
Phys.\ Rev.\ C \textbf{72}, 054310 (2005) [nucl-th/0506054]. 

\bibitem{website} The Pb nucleon configurations computed by Alvioli \textit{%
et al} can be found on the website http://users.phys.psu.edu/{\raise.17ex%
\hbox{$\scriptstyle\sim$}}malvioli/eventgenerator/ .


\bibitem{Kowalski:2003hm} H.~Kowalski and D.~Teaney, 
Phys.\ Rev.\ D \textbf{68}, 114005 (2003) [hep-ph/0304189]. 


\bibitem{Rezaeian:2012ji} A.~H.~Rezaeian, M.~Siddikov, M.~Van de Klundert
and R.~Venugopalan, 
Phys.\ Rev.\ D \textbf{87}, no. 3, 034002 (2013) [arXiv:1212.2974]. 


\bibitem{Schenke:2013dpa} B.~Schenke, P.~Tribedy and R.~Venugopalan, 
Phys.\ Rev.\ C \textbf{89}, 024901 (2014) [arXiv:1311.3636 [hep-ph]]. 


\bibitem{Teaney:2012ke} D.~Teaney and L.~Yan, 
Phys.\ Rev.\ C \textbf{86}, 044908 (2012). 


\bibitem{Niemi:2012aj} H.~Niemi, G.~S.~Denicol, H.~Holopainen and
P.~Huovinen, 
Phys.\ Rev.\ C \textbf{87}, 054901 (2013). 

\bibitem{Schenke:2010nt} B.~Schenke, S.~Jeon and C.~Gale, 
Phys.\ Rev.\ C \textbf{82}, 014903 (2010) [arXiv:1004.1408 [hep-ph]]. 

\bibitem{Marrochio:2013wla} H.~Marrochio, J.~Noronha, G.~S.~Denicol,
M.~Luzum, S.~Jeon and C.~Gale, 
arXiv:1307.6130 [nucl-th]. 

\bibitem{Gale:2012rq} C.~Gale, S.~Jeon, B.~Schenke, P.~Tribedy and
R.~Venugopalan, 
Phys.\ Rev.\ Lett.\ \textbf{110}, 012302 (2013). 

\bibitem{IS} W.~Israel and J.~M.~Stewart, Ann. Phys.\ (N.Y.) \textbf{118},
341 (1979).

\bibitem{DNMR} G.~S.~Denicol, H.~Niemi, E.~Molnar and D.~H.~Rischke, 
Phys.\ Rev.\ D \textbf{85}, 114047 (2012). 

\bibitem{Landau} L.D.\ Landau and E.M.\ Lifshitz, \textit{Fluid Mechanics},
(Pergamon; Addison-Wesley, London, U.K.; Reading, U.S.A., 1959).


\bibitem{Huovinen:2009yb} P.~Huovinen and P.~Petreczky, 
Nucl.\ Phys.\ A \textbf{837}, 26 (2010). 


\bibitem{QM14} Such fits to data were presented in the poster section by
J.-F.~Paquet \textit{et al} in the Quark Matter 2014 conference.

\bibitem{Cooper:1974mv} F.~Cooper and G.~Frye, 
Phys.\ Rev.\ D \textbf{10}, 186 (1974). 


\bibitem{Denicol:2014vaa} G.~S.~Denicol, S.~Jeon and C.~Gale, 
arXiv:1403.0962 [nucl-th]. 
\end{thebibliography}
\end{document}